\documentclass[prd,nofootinbib,showpacs,preprint]{revtex4}

\usepackage{graphicx}
\usepackage{hyperref}
\usepackage{indentfirst}
\usepackage{amssymb} 

\newcommand{\vev}[1]{\langle #1 \rangle}
\newcommand{\ket}[1]{\vert #1 \rangle}

\begin{document}

\title{On stability of false vacuum in supersymmetric theories 
with cosmic strings}

 \author{Brijesh Kumar} 
 \email{brijesh@phy.iitb.ac.in}
 \author{Urjit A. Yajnik}
 \email{yajnik@phy.iitb.ac.in}
 \affiliation{Department of Physics, Indian Institute of Technology, 
Bombay, Mumbai - 400076, India}

\begin{abstract}
We study the stability of supersymmetry breaking vacuum in the
presence of cosmic strings arising in the messenger sector.
For certain ranges of the couplings, the desired supersymmetry
breaking vacua become unstable against decay into phenomenologically
unacceptable vacua. This sets constraints on the range of allowed 
values of the coupling constants appearing in the models and
more generally on the chosen dynamics of gauge symmetry breaking.
\end{abstract}

\maketitle

\section{Introduction}

\indent Supersymmetry is a powerful principle which brings control
in Quantum Field Theory. In particular its non-renormalization
properties\cite{Salam:1974jj}\cite{Grisaru:1979wc} and its potential 
for solving the hierarchy problem 
\cite{Witten:1981nf,Dimopoulos:1981zb,Kaul:1981hi,Sakai:1981gr}
make it appealing as an ingredient of phenomenology, enabling
a description in terms of weakly coupled degrees of freedom.
In practice however, the implementation of the principle 
requires breaking supersymmetry in a hidden sector and communicating
the effects to observed phenomenology indirectly;
for reviews see \cite{Martin:1997ns,Chung:2003fi}. The issue is 
further  complicated by the fact that breaking supersymmetry 
is not itself a generic possibility. It was elucidated by
\cite{Nelson:1993nf} that presence of R parity and its
breakdown can play a crucial diagnostic role for understanding
generic possibilities of breakdown. A persistent problem
of supersymmetry breakdown however is that
such vacua may be only local minima and have the danger of
relaxing to a true minimum where supersymmetry is restored.
This unpleasant possibility is however easy to avoid if
the tunneling rate to the true vacuum can be made much longer than
the known age of the Universe.
New insights into consistency of supersymmetry breakdown in
metastable vacua have been obtained
recently in \cite{Intriligator:2006dd, Intriligator:2007py} in the
context of supersymmetric QCD. In such models several possibilities
exist, e.g., supersymmetry breaking vacuum need not be unique but 
could be degenerate with several supersymmetry preserving vacua. 
The issues involved in phenomenological implementation of these 
ideas have also been elucidated in \cite{Dine:2007dz,Banks:2006ma,
Banks:2005df,Intriligator:2007py,Intriligator:2006dd, Dine:2006gm,
Shih:2007av,Dine:2006xt,Essig:2007kh,Aharony:2006my}. Applications
to cosmology appear for example in \cite{Fischler:2006xh, Endo:2007sw}

The question of whether the metastable vacua are cosmologically stable  
against quantum tunneling for all ranges of couplings received
early attention in \cite{Dasgupta:1996pz,Craig:2006kx,Abel:2006cr}
in the context of Gauge Mediated Supersymmetry Breaking (GMSB)
(see for instance the reviews \cite{Giudice:1998bp}
\cite{Dubovsky:1999xc}).
In these models supersymmetry is dynamically broken in a hidden 
sector at a scale $\Lambda_s$ and communicated to the standard model 
through a ``messenger" sector at TeV scale.
The messenger sector has suitable interactions with both the hidden 
and visible sectors.  The true vacuum in some of these models is 
supersymmetric and color breaking. Supersymmetry is broken and 
colour is preserved only in metastable vacua.

In this paper we advance the point that it is not sufficient
to study the stability of supersymmetry breaking vacua in
their translationally invariant i.e., spatially homogeneous avatar. 
Generically topological defects such as cosmic strings or 
monopoles may obtain in the course of implementation of 
any particular scheme. Such defects could arise in the early 
universe \cite{Kibble:1980mv}. It is then necessary to take 
account of their presence when studying stability issues. 
A class of topological defects can exist
which can nucleate the formation of the true vacuum in
such a way that the exponential suppression inherent in tunneling 
phenomenon no longer obtains. Specifically, when a cosmic string is 
present in a false vacuum, such a local minimum can be rendered 
unstable against decay to a vacuum of lower 
energy \cite{Yajnik:1986wq}\cite{Dasgupta:1997kn}. 
This  process was studied by one of us \cite{Yajnik:1986tg} in the context 
of  phase  transitions in Grand Unified Theory (GUT) models. It was shown 
that generically, the presence of the cosmic string entailed the 
consequence that false vacuum would ``roll-over" smoothly to 
the true vacuum without recourse to quantum tunneling. In this 
way, a putative first-order  phase transition becomes second order,
with important cosmological implications.

The same process is applicable here in the context of gauge mediated
supersymmetry breaking models. It is possible to establish the required results
numerically where formal methods provide a suggestive answer. We study a model of 
supersymmetry breaking which contains two classes of vacua, color breaking and
color preserving. The color preserving metastable vacuum becomes parametrically
unstable in the presence of cosmic strings. This puts constraints on the model
of messenger sector being used and may make some models unviable.

This paper is organized as follows. The next section \ref{minima} provides 
a brief review of the class of models we have studied here. Section \ref{eom} discusses 
the vortex (strings) ansatz possible in the relevant vacua and the equations 
of motion to be solved. The results of the numerical solutions are 
described in section \ref{numsol}. A semi-analytic approach explaining the 
numerical results is discussed in section \ref{sec:stabans}, followed by concluding remarks 
in section \ref{conc}.

\section{A model for messenger sector}
\label{minima}

In this section, we summarize the models of \cite{Dine:1995ag, Dine:1994vc}
as discussed in \cite{Dasgupta:1996pz}. Supersymmetry is broken dynamically 
in the hidden sector. The messenger sector shares a symmetry $G_m$ with the
hidden sector. It is sufficient for $G_m$ 
to be global, but we consider the possibility of $G_m$ being local, permitting
cosmologically viable strings. The case of global $G_m$ is also commented upon.
The messenger sector is responsible for transmitting supersymmetry breaking 
from the hidden sector to the visible sector. This messenger sector consists 
of the following: a gauge singlet, $S$, a pair of messenger quarks, $q$ and 
$\overline{q}$, belonging to the $3$ and $\overline{3}$ representation of the 
$SU(3)_c$ color group, and a pair of chiral superfields, $N$ and $P$ in 
vector-like representations of $G_m$. The possibility of anomalies in the 
hidden sector forces the introduction of extra superfields $E_i$ to cancel the 
anomalies. However, in this paper, we consider only `Minimal' models which do 
not include these extra fields. The most general superpotential for the messenger 
sector is given by \cite{Dasgupta:1996pz}
\begin{equation}
W_{mes} = \kappa S \overline{q} q + \frac{\lambda}{3} S^3 + 
\lambda_1 P N S + W_1(P,N,E_i).
\end{equation}
The three coupling constants $\kappa$, $\lambda$, and $\lambda_1$ are
all positive by a suitable phase definition of the fields. In the case 
where $G_m$ is $U(1)$, if $e$ is the $U(1)$ gauge coupling and 
$P$, $N$, and $E_i$ have charges $+1$, $-1$, and $y_i$, the effective 
potential for the charged scalars after integrating out the hidden sector 
is given by \cite{Dasgupta:1996pz}
\begin{equation}
V_{SB} = M_1^2 (\vert P \vert^2 - \vert N \vert^2 + y_i \vert E_i \vert^2)
+ M_2^2 (\vert P \vert^2 + \vert N \vert^2 + y_i^2 \vert E_i \vert^2) + \dots
\end{equation}
in which the relation between the mass parameters $M_i$, $i = 1,2$ and 
the supersymmetry breaking scale $\Lambda_s$ is
\begin{equation}
M_i^2 = c_i \Lambda_s^2 (\frac{e^2}{(4\pi)^2})^i.
\end{equation}
The factors $c_i$ $(i = 1,2)$ are of order unity, with their signs dependent 
on the content of the dynamical symmetry breaking sector. There are 
higher-dimensional
terms coming from more loops and these are what the ellipsis in the 
expression for $V_{SB}$ stand for. In addition, there are contributions 
to the scalar potential coming from the $U(1)$ D-term and F-terms.
In what follows, we will consider the set of minimal models which satisfy
$ \frac{\partial W_1}{\partial P} = \frac{\partial W_1}{\partial N} = 0$.
This set includes models in which there are no E fields, models in which
$W_1 = 0$, and models in which the E fields do not couple to P and N.
After including the $U(1)$ D-terms and F-terms and setting $W_{1} = 0$, 
we consider an explicit model in which ($M_{1} = 0$), and the
scalar potential of the messenger sector becomes \cite{Dasgupta:1996pz}
\begin{eqnarray}
\label{pot}
V_{mes} &=& \frac{e^2}{2}(\vert P \vert^2 - \vert N \vert^2)^2 + (M_2^2 +
\lambda_1^2 \vert S \vert^2)( \vert P \vert^2
+ \vert N \vert^2) 
\nonumber \\ &&
+ ~\kappa^2 \vert S \vert^2(\vert q \vert^2 + \vert
\overline{q} \vert^2) + \vert \kappa \overline{q}q
+ \lambda S^2 + \lambda_1 PN \vert^2.
\end{eqnarray} 
We shall be interested in the case when $M_2^2<0$ which signals the 
breakdown of the $U(1)$. This makes $V_{mes}$ unbounded 
from below, but in the total potential, higher-order terms in $V_{SB}$
result in a deep global minimum far away in field space in which the
visible sector is supersymmetric. For a viable local minimum, one must set
$q = \overline{q} = 0$ in the expression for $V_{mes}$ and in order to
have $S \neq 0$ simultaneously, we must have
\begin{equation}
\lambda > \lambda_1.
\end{equation} 
The local minima lie at $q = \overline{q} = 0$ and
\begin{equation}
|P|^2 = |N|^2 = -M_2^2\frac{\lambda}{\lambda_1^3(2-\lambda_1/\lambda)}
\label{PN}
\end{equation} 
\begin{equation}
|S|^2 = -M_2^2\frac{1-\lambda_1/\lambda}{\lambda_1^2(2-\lambda_1/\lambda)}
\label{S}
\end{equation} 
\begin{equation}
Arg(PNS^{*2}) = \pi
\label{eq:PNSphase}
\end{equation} 
Along with the condition $\lambda > \lambda_1$, stability of the local
minima require the following relations between the couplings
\cite{Dasgupta:1996pz}:
\begin{equation}
\lambda_1^3 \leq 2\lambda e^2
\end{equation} 
\begin{equation}
\lambda_1 \leq
\frac{\kappa\lambda}{\kappa + \lambda}
\label{condition}
\end{equation} 
Condition (\ref{condition}) is also mentioned in \cite{ArkaniHamed:1996xm} 
wherein it is noted that there is still a possibility of color breaking
vacua when (\ref{condition}) holds.

The scalar potential of the messenger sector given in equation (\ref{pot}) 
contains two important types of 
vacua. One of these, which will be denoted as $\ket{V_{1}}$ later in the paper, is the 
supersymmetry breaking local minimum described above. This vacuum has $\vev{S} \neq 0$
and $\vev{q} = \vev{\overline{q}} = 0$, which means that SUSY is broken while 
the color gauge group is unbroken. The other vacuum, which shall be denoted
$\ket{V_{2}}$, also has $\vev{S} \neq 0$ but the fields $q$ and $\overline{q}$ get
non-zero VEVs. This means that $\ket{V_{2}}$ is a phenomenologically undesirable
minimum in which SUSY is still broken but the color gauge group is also broken.
If cosmic strings are supported, these vacua are modified to what will be denoted
as $\ket{V_{1}^{(string)}}$ and $\ket{V_{2}^{(string)}}$ respectively. It
will be shown in a later section that $\ket{V_{1}^{(string)}}$ becomes parametrically 
unstable towards decay into $\ket{V_{2}^{(string)}}$. The next section describes the 
available string solutions.

\section{The string ansatz} 
\label{eom}
With the scalar potential for the messenger sector as given in (\ref{pot}), 
the full Lagrangian for all fields $(P,N,S,q,\overline{q})$ can be written 
down. The fact that string solutions are cylindrically symmetric makes it
convenient to use cylindrical coordinates. We ignore the z-dependence and
look for time-independent solutions. In a local minimum, $|P| = |N|$ from
(\ref{PN}) and $\vev{S}$ is as given in (\ref{S}). The ansatz functions 
for the scalar fields are a simple generalization of the 
Abrikosov-Nielsen-Olesen string
\cite{Abrikosov:1956sx}\cite{Nielsen:1973cs},
(see also \cite{Hindmarsh:1994re})
\begin{equation}
P = \eta f(r)e^{i\theta}
\label{Pansatz}
\end{equation}
\begin{equation}
N = \eta f(r)e^{-i\theta}
\label{Nansatz}
\end{equation}
\begin{equation}
S =  \eta g(r)e^{i\pi/2}
\label{Sansatz}
\end{equation}
\begin{equation}
q = \overline{q} = \eta h(r)e^{i\pi/2}
\label{qansatz}
\end{equation}
The value of $\eta$ is given by the right hand side of equation (\ref{PN}) 
Note that the winding number for the P and N fields has been
chosen equal and opposite. Single valuedness requires that the
phase complete a circuit of $2\pi$ at infinity. This requires 
the phase to be an integer multiple of cylindrical angle $\theta$. 
Along with the choice of phase for $S$ as in (\ref{Sansatz}), this 
winding ensures the condition (\ref{eq:PNSphase}). This is the 
expected minimal energy configuration.
The phases for $q$, and $\overline{q}$ satisfy the expectation
that their product remains negative. At the core of the vortex, the fields
P and N must vanish, and hence $f(0) = 0$. At infinity, P and N approach their 
vacuum expectation values which are denoted by $\eta$, and hence 
$f(\infty) = 1$. Similarly, $g(r)$ approaches the value 
given in (\ref{S}) at infinity but can be non-zero in the core. The value of 
$h(r)$ can be non-zero in the core also, but its value at infinity is of 
importance in determining whether the resulting vacuum is phenomenologically 
viable or not. This is because the vacuum with $h$ and hence $q$ and 
$\overline{q}$ non-zero is color-breaking.

A vortex solution in P and N also includes a coupling of the fields to a
gauge field $A(r)$. The gauge field behavior is described by a
function $a(r)$ defined through
\begin{equation}
A_{\theta}(r) = \frac{1}{er}a(r)
\label{aansatz}
\end{equation}
\begin{equation}
A_{0} = A_{r} = A_{z} = 0
\end{equation}
where for continuity of $A_{\theta}(r)$, we have $a(0)=0$, and e is the
unit of the abelian $G_m$ charge. At infinity, $A_{\theta}(r)$ is 
pure gauge and goes as $1/r$ 
and hence $a(\infty)=1$. The Lagrangian for the system can be written as:
\begin{equation}
 L = \frac{1}{2}(\partial_{\mu} S)^2 + \frac{1}{2}(\partial_{\mu} q)^2 + 
\frac{1}{2}(\partial_{\mu} \overline{q})^2 + \frac{1}{2}\vert (\partial_{\mu} + 
ieA_{\mu})P \vert^2 + \frac{1}{2}\vert (\partial_{\mu} + 
ieA_{\mu})N \vert^2 - \frac{1}{4}F_{\mu\nu}F^{\mu\nu} - V_{mes}
\end{equation}
where $V_{mes}$ is as given in (\ref{pot}).
After substituting equations (\ref{Pansatz}\,-\,\ref{aansatz}) into 
the Lagrangian, the equations of motion 
can be written in terms of $f(r)$, $a(r)$, $g(r)$, and $h(r)$. They are a 
set of coupled second-order and non-linear ordinary differential equations 
given by:
\begin{equation}
\frac{d^2f}{dr^2} + \frac{1}{r}\frac{df}{dr} - \frac{1}{r^2}f(a-1)^2 
-\left[ \frac{M_2^2}{\eta^2} - \lambda_1(\lambda - \lambda_1)g^2 -
\lambda_1\kappa h^2 \right]f - \lambda_1^2f^3 = 0
\label{feom}
\end{equation} 
\begin{equation}
\frac{d^2a}{dr^2} - \frac{1}{r}\frac{da}{dr} - 2e^2f^2(a-1) = 0
\label{aeom}
\end{equation} 
\begin{equation}
\frac{d^2g}{dr^2} + \frac{1}{r}\frac{dg}{dr} -\left[ 2\lambda_1(\lambda_1 -
\lambda)f^2 + 2\kappa(\kappa + \lambda)h^2 \right]g - 2\lambda^2g^3 = 0
\label{geom}
\end{equation} 
\begin{equation}
\frac{d^2h}{dr^2} + \frac{1}{r}\frac{dh}{dr} - \left[ \kappa(\kappa +
\lambda)g^2 - \kappa\lambda_1f^2 \right]h - k^2h^3 = 0
\label{heom}
\end{equation} 

The fields P and N in which we have setup string configurations have a
$U(1)$ symmetry. As is well known, existence of cosmic string solutions
requires that the space of equivalent vacua be necessarily multiply
connected. Generically in a larger gauge group such as $SO(N)$, 
$Spin(N)$ or $SU(N)$, cosmic strings are in principle possible for $P$ and
$N$ in any representation other than the vector or the fundamental.
Thus, the conclusions of this paper have relevance to more general cases.

\section{Testing the vacua}
\label{numsol}

\subsection{Homogeneous vacuum}
\label{sec:homog}
We begin with calculating the field values at infinity which can be obtained 
easily from equations (\ref{feom}\,-\,\ref{heom}). At $r=\infty$, all the 
derivative terms vanish due to translational invariance and so do the terms 
containing $(1/r^2)$. The equations get reduced to the following 
simple set of simultaneous equations:
\begin{equation}
\left[ \frac{M_2^2}{\eta^2} - \lambda_1(\lambda - \lambda_1)g^2 -
\lambda_1\kappa h^2 \right]f + \lambda_1^2f^3 = 0
\label{vac1}
\end{equation} 
\begin{equation}
 2e^2f^2(a-1) = 0
\label{vac2}
\end{equation} 
\begin{equation}
\left[ 2\lambda_1(\lambda_1 -
\lambda)f^2 + 2\kappa(\kappa + \lambda)h^2 \right]g + 2\lambda^2g^3 = 0
\label{vac3}
\end{equation} 
\begin{equation}
\left[ \kappa(\kappa +
\lambda)g^2 - \kappa\lambda_1f^2 \right]h + k^2h^3 = 0
\label{vac4}
\end{equation} 
One must note that values for three couplings $(\kappa, \lambda, \lambda_1)$ 
must be given for each unique set of vacuum solutions. The set of vacuum 
solutions consists of four solutions. 
One is the trivial solution in which all fields vanish. Another is one 
in which $f$ and $a$ are non-zero but SUSY breaking is not being communicated since 
$g$ and $h$ are both zero. 

The other two translation invariant minima shall be denoted $\ket{V_{1}}$ and 
$\ket{V_{2}}$. Their field content is as follows:
\begin{equation}
\ket{V_{1}}: f = a = 1 ~~~g \neq 0 ~~~h=0
\label{vacuum1}
\end{equation} 
\begin{equation}
\ket{V_{2}}: f \lesssim 1 ~~~a = 1 ~~~g \neq 0 ~~~h \neq 0
\label{vacuum2}
\end{equation}
$\ket{V_{1}}$ is the desired supersymmetry breaking and color preserving local
minimum described in section 2. The value of g is exactly as required by
equation (\ref{S}), and it is a function of the couplings. $\ket{V_{2}}$ is an
undesirable minimum in which $h$ and hence $q$ and $\overline{q}$ get VEVs.
Furthermore, what is interesting is that for all ranges of couplings, its
energy is always lower than that of $\ket{V_{1}}$. 
\begin{equation}
\vev{V_{2} \vert V_{mes} \vert V_{2}} < \vev{V_{1} \vert V_{mes} \vert V_{1}}
\end{equation}
The expression for
$V_{mes}$ in terms of $(f,g,h)$ can be written as
\begin{equation}
V_{mes} = \eta^4 \left[ 2f^2\left(\lambda_1^2g^2 - 
\frac{\vert M_{2} \vert^2}{\eta^2}\right)
+ 2\kappa^2g^2h^2 + (\lambda_1f^2 - \lambda g^2 - \kappa h^2)^2 \right]
\label{fghequation}
\end{equation}
It is a simple exercise to check that for each set of solutions, $V_{mes}$
is always smaller for $\ket{V_{2}}$ compared to $\ket{V_{1}}$.

\subsection{Vacuum with gauge string}
\label{sec:vacggestr}

\begin{figure}[!htp]
\begin{center}
\includegraphics[width=0.8\textwidth]{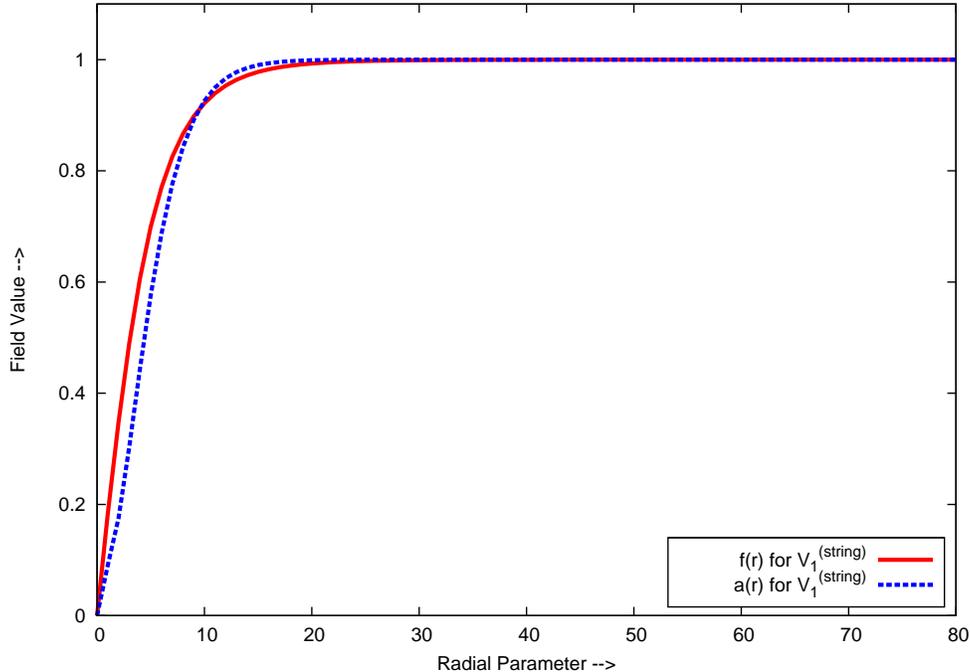} 
\caption{f(r) and a(r) for $\ket{V_{1}^{(string)}}$ with boundary conditions as
in eqn. (\ref{vacuum1}), and $\kappa = 1.3, \lambda = 1.45,
\lambda_{1}=0.65$. Here, $f_{\infty} = 1$ and $a_{\infty} = 1$.}
\label{fig1}
\end{center}
\end{figure}

A numerical strategy exists for solving equations 
(\ref{Pansatz}\,-\,\ref{aansatz}).
This consists of using relaxation techniques, after discretizing the
equations of motion on a grid converting them to a set of coupled
polynomial equations. In conformity with physical expectations
the initial guess is one with $f=a=0$ in the 
core and with $g$ and $h$ arbitrary. At
infinity, the trial is chosen to be either settling into $\ket{V_{1}}$ or
$\ket{V_{2}}$. The exact values of $(f,a,g,h)$ in $\ket{V_{1}}$ and
$\ket{V_{2}}$ can be found by first solving equations (\ref{vac1}\,-\,\ref{vac4}).
We denote by $\ket{V_{1}^{(string)}}$
the static state of the system in which the cosmic string configuration
approaches $\ket{V_{1}}$ (upto a $\theta$-dependent phase) as $\vert\vec{r}\,\vert \rightarrow \infty$. 
Similarly, $\ket{V_{2}^{(string)}}$ denotes a static string solution which asymptotes to 
$\ket{V_{2}}$. An example of the solutions in a case where both 
$\ket{V_{1}^{(string)}}$ and $\ket{V_{2}^{(string)}}$ are stable is shown in 
figures \ref{fig1}\,-\,\ref{fig4}. 

\begin{figure}[!htp]
\begin{center}
\includegraphics[width=0.8\textwidth]{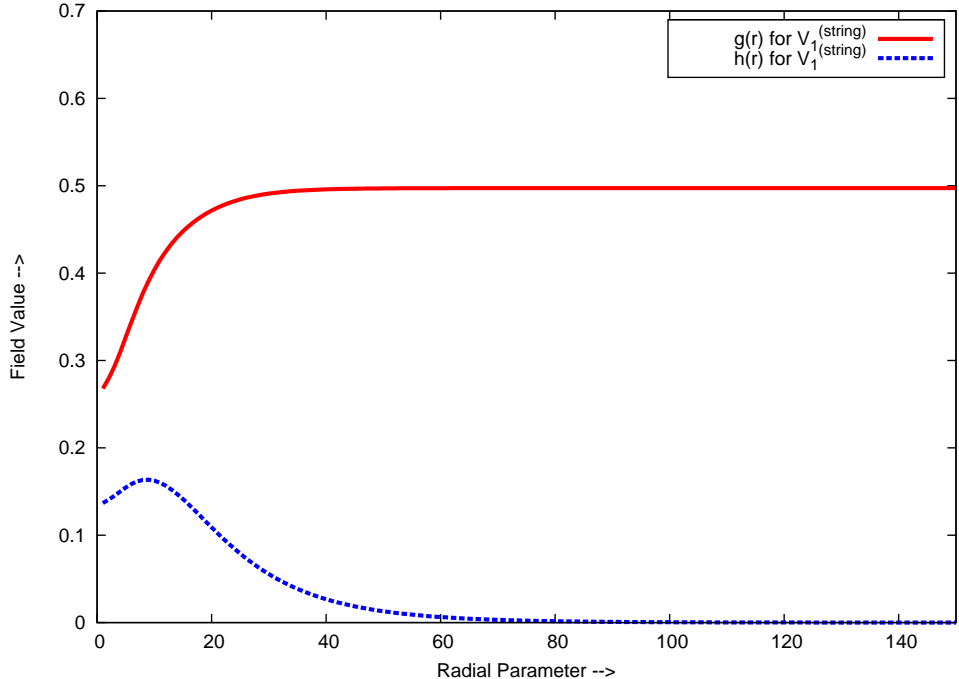} 
\caption{g(r) and h(r) for $\ket{V_{1}^{(string)}}$ with boundary conditions as
in eqn. (\ref{vacuum1}), and $\kappa = 1.3, \lambda = 1.45,
\lambda_{1}=0.65$. Here, $g_{\infty} = 0.497317$ and $h_{\infty} = 0$.}
\label{fig2}
\end{center}
\end{figure}

\begin{figure}[!htp]
\begin{center}
\includegraphics[width=0.8\textwidth]{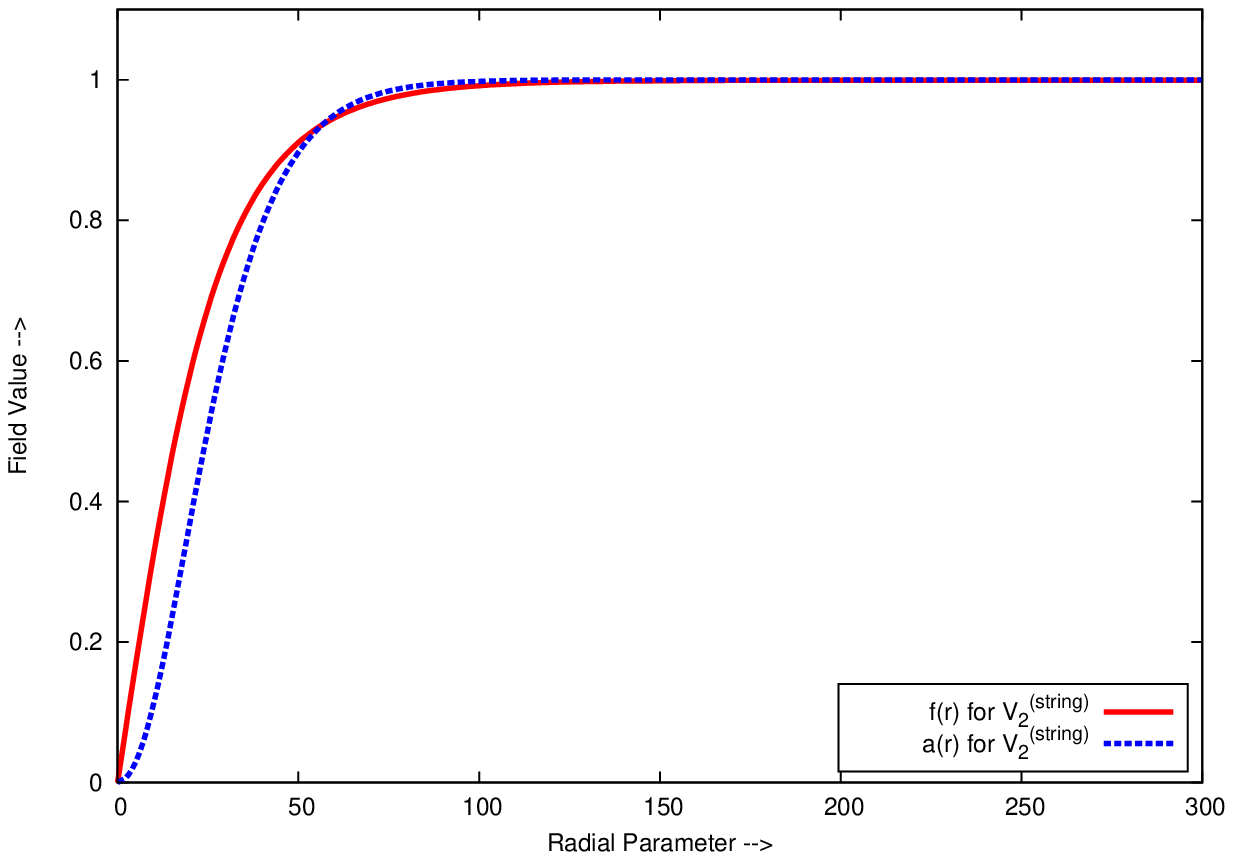} 
\caption{f(r) and a(r) for $\ket{V_{2}^{(string)}}$ with boundary conditions as
in eqn. (\ref{vacuum2}), and $\kappa = 1.3, \lambda = 1.45,
\lambda_{1}=0.65$. Here, $f_{\infty} = 0.999405 $ and $a_{\infty} = 1$.}
\label{fig3}
\end{center}
\end{figure}

\begin{figure}[!htp]
\begin{center}
\includegraphics[width=0.8\textwidth]{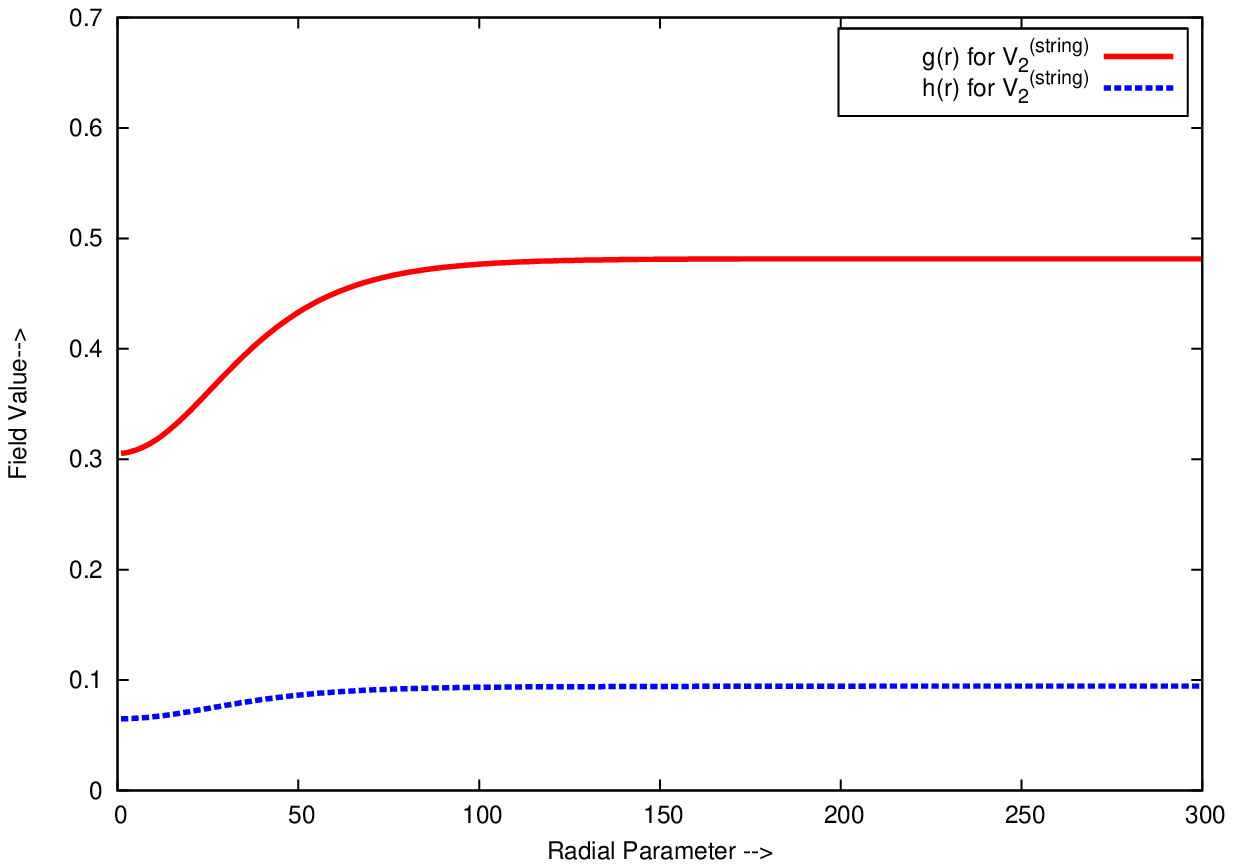} 
\caption{g(r) and h(r) for $\ket{V_{2}^{(string)}}$ with boundary conditions as
in eqn. (\ref{vacuum2}), and $\kappa = 1.3, \lambda = 1.45,
\lambda_{1}=0.65$. Here, $g_{\infty} = 0.481526$ and $h_{\infty} = 0.094434$.}
\label{fig4}
\end{center}
\end{figure}

It may happen that for some values 
of the couplings, a solution with the expected field values at infinity cannot 
be obtained. When this happens, that solution with its asymptotic vacuum 
values is not tenable. We have explored the stability of $\ket{V_{1}^{(string)}}$ 
and $\ket{V_{2}^{(string)}}$ over  a range of 
the couplings. For the entire range of couplings studied, the string 
solution $\ket{V_{2}^{(string)}}$ is always obtained. However,
no stable static solution $\ket{V_{1}^{(string)}}$ 
can be obtained for certain ranges which are indicated in table \ref{table1}. 
Note specifically that for $\kappa = 1$ and $\lambda = 1.1$,
values of $\lambda_{1} < 0.1$ which are of phenomenological interest are
immediately ruled out. It is for these ranges of the couplings that the 
corresponding metastable
vacua are no longer local minima due to the presence of cosmic strings.

In the context of the early universe, the parameters in the effective action are
as a rule temperature dependent. The above results then imply that we may start
with a phase wherein a state of the type $\ket{V_{1}^{(string)}}$ may be a local
minimum but with reduction of temperature, the string solution may not exist in 
the sense argued above. How this comes about in the present example
is taken up at the end of sec. \ref{sec:stabans}. The subsequent dynamics 
can be simulated by restoring
time dependence in equations (\ref{feom}\,-\,\ref{heom}). It was shown in 
\cite{Yajnik:1986tg} that the string configuration rendered parametrically
unstable undergoes a real time evolution into a possible stable string 
configuration with a modified vacuum at infinity. This can be referred to as
a ``roll-over'' process, a prompt, semi-classical evolution rather than
tunneling by spontaneous formation of bubbles.

\begin{table}[!htp]
\begin{center}
\begin{tabular}{|c|c|c|c|c|}
\hline
 $\kappa$ & $\lambda_1$ & $\lambda$ & $\ket{V_{1}^{(string)}}$ & $\ket{V_{2}^{(string)}}$ \\
\hline \hline
 2.4 & 1.2 & $\geq 2.41$ & $\times$ & $\checkmark$ \\
 \hline
 2 & 1 & $\geq 2.01$ & $\times$ & $\checkmark$ \\
 2 & $\leq 0.75$ & 2.5  & $\times$ & $\checkmark$ \\
 \hline
 1.7 & 0.85 & $\geq 1.71$ & $\times$ & $\checkmark$ \\
 \hline
 1.3 & 0.65 & $\leq 1.42$  & $\times$ & $\checkmark$ \\
 1.3 & 0.65 & $\geq 1.43$  & $\checkmark$ & $\checkmark$ \\
 1.3 & $\leq 0.65$ & 1.5 & $\checkmark$ & $\checkmark$ \\
 \hline
 1 & 0.5 & $\leq 1.07$ & $\times$ & $\checkmark$ \\
 1 & 0.5 & $\geq 1.1$  & $\checkmark$ & $\checkmark$ \\
 1 & $\geq 0.3$ & 1.1 & $\checkmark$ & $\checkmark$ \\
 1 & $\leq 0.15$ & 1.1 & $\times$ & $\checkmark$ \\
 \hline
 0.6 & 0.3 & $\leq 0.68$ & $\times$ & $\checkmark$ \\
 \hline
\end{tabular}
\caption{Testing the existence of local minima viable in the presence of a
cosmic string. Tick marks indicate when solutions are found and crosses 
indicate when solutions cannot be obtained. Vacua of type 1, eqn. (\ref{vacuum1}),
become parametrically disallowed due to the presence of a cosmic string.}
\label{table1}
\end{center}
\end{table}

\subsection{Global cosmic strings}
\label{sec:vacglbstr}

\begin{table}[!htp]
\begin{center}
\begin{tabular}{|c|c|c|c|c|}
\hline
 $\kappa$ & $\lambda_1$ & $\lambda$ & $\ket{V_{1}^{(string)}}$ & $\ket{V_{2}^{(string)}}$ \\
\hline \hline
 1.3 & 0.65 & $\geq 1.3$ & $\times$ & $\checkmark$ \\
\hline
 1 & 0.5 & $\leq 1.19$ & $\times$ & $\checkmark$ \\
 1 & 0.5 & $\geq 1.20$ & $\checkmark$ & $\checkmark$ \\
\hline
\end{tabular}
\caption{Sample results for global strings, i.e., no $a(r)$ field present.
These are to be compared to rows 5\,-\,9 of table \ref{table1}.}
\label{table2}
\end{center}
\end{table}

Our demonstration though specific to messenger group of this example
may have relevance to a more general setting where some sector of the
theory possesses global symmetries which are broken in the
desirable vacuum. In
this case, the strings arising are global cosmic strings. In our example 
this corresponds to setting $a(r)$ identically to zero. We have looked for 
solutions without the $a(r)$ field for some ranges of the couplings. The
results are shown in table \ref{table2}. It is found that this condition rules 
out $\ket{V_{1}^{(string)}}$ for a larger range of couplings as compared to the 
case in which $a(r)$ is present. The results for $\ket{V_{2}^{(string)}}$ remain 
unchanged.

A global cosmic string network is not desirable for stable cosmology \cite{Kibble:1980mv}.
This is because the energy per unit length of a global string is 
divergent and can dominate the energy density of the Universe.
However the divergence of the energy  with distance from the
string is logarithmic and a transient network of such strings
cannot be ruled out. We expect such a network to eventually
relax to a homogeneous vacuum. Even a transient network however,
may be capable of destabilizing a particular local minimum
according to the discussion of this subsection.

\section{Stability Analysis of $\ket{V_{1}^{(string)}}$}
\label{sec:stabans}

In the previous section, the stability of $\ket{V_{1}^{(string)}}$ has been studied
numerically. As illustrated in table \ref{table1}, there are certain domains 
in the parameter space of $\kappa$, $\lambda_1$, and $\lambda$ in which 
$\ket{V_{1}^{(string)}}$ is not admissible. The non-existence of such solutions was then
understood as an unavailability of $\ket{V_{1}^{(string)}}$ for those parameter values.
This numerical result can be confirmed by a semi-analytic 
treatment following an approach discussed in \cite{Yajnik:1986wq}. 
 
A preliminary analysis can be made for the translation invariant vacuum to study the
effect of changing $\lambda$. Referring to equation (\ref{fghequation}), the effective 
squared mass $m_{eff}^2$ for the $h$ field is given by
\begin{equation}
 m_{eff}^2 = 2g^2(\kappa^2 + \kappa\lambda) - 2\kappa\lambda_{1}f^2
\label{eq:heffmass}
\end{equation}
Now if $\kappa$ and $\lambda_{1}$ are held fixed, when $\lambda$ is large enough, this
quantity is positive and $h=0$ is admissible as a local minimum. As $\lambda$ reduces, 
there comes a point when $m_{eff}^2$ becomes negative. This corresponds to the regions of 
instability shown in table \ref{table1}. For example, when $\kappa = 1.3$ and 
$\lambda_{1}$ = 0.65, reducing $\lambda$ below 1.43 drives its effective squared mass 
negative making $h=0$ unstable. 

We now proceed along the lines of \cite{Yajnik:1986wq} to analyze the linear stability of 
$\ket{V_{1}^{(string)}}$ by adding a small time-dependent term to the time-independent 
numerical solution. If small oscillations around this solution have only real 
frequencies, the configuration is considered stable. For a nonavailability of 
$\ket{V_{1}^{(string)}}$, at least one of the modes of oscillation must possess an 
imaginary frequency. The stability analysis for $\ket{V_{1}^{(string)}}$ is greatly 
simplified by the following
observation. The fields $f$ and $g$ do not differ significantly in states $\ket{V_{1}^{(string)}}$ 
and $\ket{V_{2}^{(string)}}$ as seen from figures \ref{fig1}\,-\,\ref{fig4}. 
Only $h$ differs significantly and the possible time-dependence of the background
fields $f$ and $g$ can be ignored. The equation of motion for $h$ with time-dependence 
restored is
\begin{equation}
-\frac{d^2h}{dt^2} + \frac{d^2h}{dr^2} + \frac{1}{r}\frac{dh}{dr} - \left[ \kappa(\kappa +
\lambda)g^2 - \kappa\lambda_1f^2 \right]h - k^2h^3 = 0
\label{heom2}
\end{equation}
and we now write
\begin{equation}
h(r,t) = \tilde{h}(r) + p(r)e^{i\omega t}
\label{timedeph}
\end{equation}
Here, $\tilde{h}$ is the time-independent solution obtained for the h field and 
$p << \tilde{h}$. Similarly, the time-independent fields $f$ and $g$ in equation 
(\ref{heom2}) are written as $\tilde{f}$ and $\tilde{g}$. Substituting equation 
(\ref{timedeph}) in (\ref{heom2}) and linearizing the equation for $p(r)$, we get 
\begin{equation}
 \omega^2 p = -\lbrack \frac{d^2}{dr^2} + \frac{1}{r}\frac{d}{dr} \rbrack \,p 
+ \lbrack 3\kappa^2 \tilde{h}^2 + \kappa (\kappa + \lambda)\tilde{g}^2 -
\kappa \lambda_{1}\tilde{f}^2 \rbrack \,p
\label{equivsch}
\end{equation}
Equation (\ref{equivsch}) has
the form of a one-dimensional Schrodinger equation with a potential $u(r)$ given by
\begin{equation}
 u(r) = 3\kappa^2 \tilde{h}(r)^2 + \kappa (\kappa + \lambda)\tilde{g}(r)^2 -
\kappa \lambda_{1}\tilde{f}(r)^2
\label{equivpot}
\end{equation}

\begin{figure}[!htp]
\begin{center}
\includegraphics[width=0.8\textwidth]{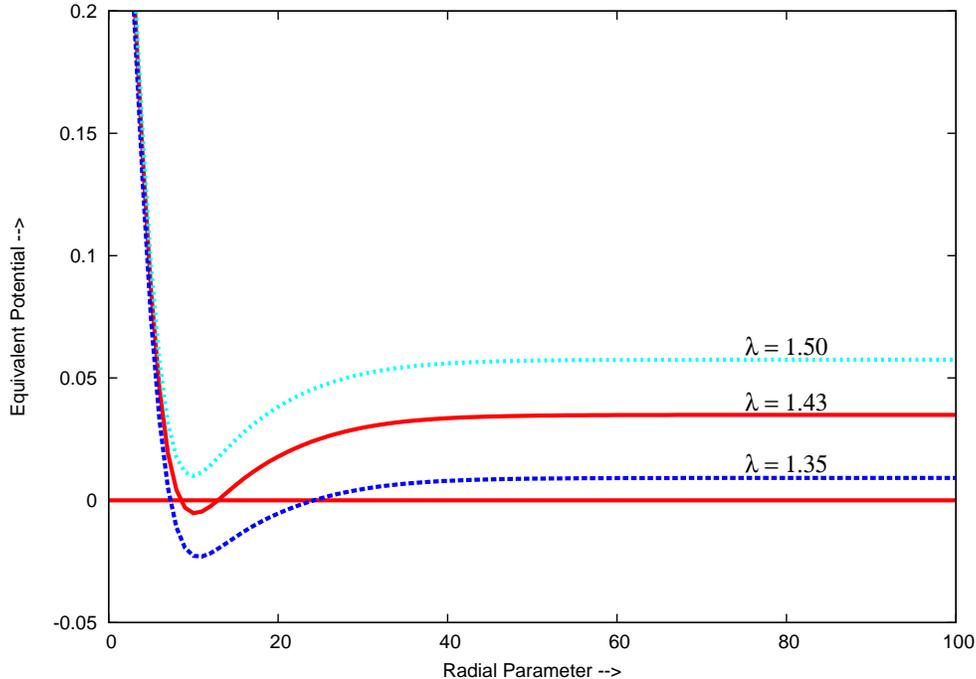} 
\caption{The equivalent Schrodinger potential of equation (\ref{equivsch}) with $\kappa = 1.3, 
\lambda_{1}=0.65$, and different values of $\lambda$. A positive energy minimum results
in a stable solution. When a negative energy bound state is possible, vacua of type 
$\ket{V_{1}^{(string)}}$ become unavailable.}
\label{fig5}
\end{center}
\end{figure}

Looking for imaginary modes of the frequency $\omega$ now reduces to finding negative-energy
bound states for this potential. Figure \ref{fig5} depicts the behavior of the potential
for values of couplings which lie in both the stable and unstable regions for
$\ket{V_{1}^{(string)}}$. Referring to table \ref{table1}, when $\kappa = 1.3$ and 
$\lambda_{1}=0.65$, $\ket{V_{1}^{(string)}}$ is stable for $\lambda = 1.5$. This is in agreement 
with the fact that the minimum of the potential for $\lambda = 1.5$ has positive energy. 
As $\lambda$ reduces towards $\lambda = 1.43$, notice that the minimum of the energy starts 
reducing until it starts becoming negative near $\lambda = 1.43$. This once again confirms 
the numerical result that vacua of type $\ket{V_{1}^{(string)}}$ become unavailable for values
of $\lambda < 1.43$.
We have similarly computed the potential $u(r)$ for critical values of $\kappa$ and $\lambda$, 
and found similar solutions confirming the results stated in table \ref{table1}.

Returning to our comment on the early Universe setting at the end of sec. \ref{sec:vacggestr}, 
the field $f$ (the VEV of $P$ and $N$) has negative effective
mass-squared, $M_2^2<0$. High temperature corrections from gauge sector add a 
term of the form $A T^2$ with $A>0$ to this effective mass-squared \cite{Weinberg:1973ua, Dolan:1973qd},
which would drive the temperature dependent value $f_T$ to zero at high temperature, restoring 
the gauge symmetry. Below the symmetry breaking critical temperature $f_T$ becomes non-zero. 
As long as this value remains small, the effective mass-squared for the $h$ 
in eq. (\ref{eq:heffmass}) remains positive. This makes $\ket{V_{1}}$ a valid local 
minimum and admits a vortex sector $\ket{V_{1}^{(string)}}$. At an even lower temperature 
$f_T$ grows sufficiently large, and the roll-over transition sets in.

\section{Conclusion}
\label{conc}

One of the main requirements in the minimal models of gauge mediated
supersymmetry breaking is that the messenger squarks do not develop VEVs
\cite{Dasgupta:1996pz}. In this paper, we have shown an example in which cosmic strings 
are present in the metastable vacua. These configurations can roll-over 
classically into undesirable configurations which asymptote into vacua in 
which $\vev{q} = \vev{\overline{q}} \neq 0$. This happens for certain ranges 
of the couplings $\kappa$, $\lambda$ and $\lambda_1$, including the 
phenomenologically interesting region in which $\lambda_{1}< 0.1$. For a
general messenger group $G_{m}$ stable cosmic strings would exist depending
on the choice of representation of the fields P and N. Furthermore, we have
evidence that if $G_{m}$ is global the instability issue becomes more
severe. In all such models, it should be possible to obtain new constraints 
on the coupling constants from detailed simulations of the cosmic strings.
Such studies provide tighter constraints on the parameters as compared to 
those from quantum tunneling effects. 

In principle, these techniques are applicable to the SUSY breaking sector itself
provided the model supports solitonic solutions. This aspect of the problem will 
be dealt with in future works.

\section{Acknowledgment}
We thank P. Ramadevi for fruitful discussions and comments. This work was
completed while the authors were at Indian Institute of Technology, Gandhinagar,
whose hospitality is acknowledged. 



\end{document}